\UseRawInputEncoding
\documentclass[aps,prd,onecolumn,groupedaddress,nofootinbib,amssymb]{revtex4}
\usepackage[dvips]{graphicx,color}
\usepackage{amssymb}
\usepackage{amsmath}
\usepackage{graphicx}
\usepackage{amsfonts}

\newcommand{\be}{\begin{equation}}
\newcommand{\ee}{\end{equation}}
\newcommand{\bea}{\begin{eqnarray}}
\newcommand{\eea}{\end{eqnarray}}
\newcommand{\beaa}{\begin{eqnarray*}}
\newcommand{\eeaa}{\end{eqnarray*}}

\begin{document}


\title{\bf D-bound and Bekenstein Bound for
the Surrounded Vaidya Black Hole}
\author{H. Hadi$^{a}$ }\email{hamedhadi1388@gmail.com}
\author{F. Darabi$^{a}$}\email{f.darabi@azaruniv.ac.ir}
\author{K. Atazadeh$^{a}$ }\email{atazadeh@azaruniv.ac.ir}
\author{Y. Heydarzade$^{b}$ }\email{yheydarzade@bilkent.edu.tr}
\affiliation{$^{a}$ Department of Physics, Azarbaijan Shahid Madani University, 53714-161 Tabriz, Iran\\
$^{b}$ Department of Mathematics, Faculty of sciences, Bilkent University, 06800 Ankara, Turkey}
\date{\today}

\begin{abstract}

We study the Vaidya black hole surrounded by
 the exotic quintessence-like, phantom-like
and cosmological constant-like fields by means of entropic
considerations. Explicitly, we show that for this thermodynamical system,
the   requirement for the identification of D-bound
and  Bekenstein entropy bound  can be considered as a thermodynamical criterion by which one can rule out the quintessence-like and phantom-like
fields, and prefer the cosmological
constant  as a viþable cosmological field.

 \end{abstract}
\maketitle

\section{Introduction}
The simplest candidate for dark energy in the context of accelerating universe
is the cosmological constant. The relevant cosmological model including  the cosmological constant is so called $\Lambda$ Cold Dark Matter $(\Lambda$CDM) model
which is consistent with current observations. However,  $\Lambda$CDM  suffers from
two well known problems, namely the coincidence problem \cite{CP} and cosmological constant problem \cite{CCP}.
To resolve these problems,  some exotic models of dark energy,  like Quintessence  and Phantom fields have been introduced. The quintessence field with a dynamical
equation of state,  having the negative equation of state parameter  $-1<\omega_{q}<-\frac{1}{3}$,
is capable of describing the late-time cosmic acceleration \cite{Caldwell}. Quintessence field plays an important role in the cosmological dynamics including matter and radiation \cite{fujii,ford,wett,ratra}.  The phantom field, as
 another candidate
for dark energy, has also a negative  equation of state parameter  $\omega_{p}<-1$
capable of describing the current acceleration of the universe, as well
\cite{carroll12,singh12,sami12}.  In the limit of  approaching  $\omega_{p}$
to a constant value,
 a big-rip singularity is resulted as a new problem. Having  these three models of dark energy successful in
predicting an accelerating cosmic dynamics does not imply  their   perfect description of current accelerating universe and one is highly motivated to revisit these three
dark energy models from a {\it non-cosmic} point of view.

 The powerful thermodynamical approach and the relevant entropic limits
can be considered as such non-cosmic point of view in the study of dark energy models. In principle, the equations of motion can perfectly   predict
the dynamical behavior of  time-reversible physical systems, however   in reality, for thermodynamical systems
the time-reversibility  is not observed  because of the entropic
consideration.  Dynamical black holes, surrounded by the cosmological fields, are such  relevant examples of  thermodynamical system, in the present study. Explicitly, we  impose an entropic criterion on the  Vaidya black hole surrounded by
some exotic fields possessing an average equation of state like the quintessence,
phantom and cosmological constant. By means of this criterion, we investigate which of these fields  can be
singled out as  the most viable cosmological field. 

In the 70s of last century, the quantum physics of black holes started by
the works of Bekenstein \cite{bekenstein3,moon1} and Hawking \cite{hawking1}.
There is a general conviction that Hawking radiation \cite{hawking1} and Bekenstein-Hawking
entropy \cite{bekenstein3,hawking1}  are the main features of a yet unknown theory of quantum gravity
which will be able to unify Einstein's general theory of relativity with
quantum mechanics. In fact, some of the experts on quantum gravity claim that black holes are the fundamental bricks of quantum gravity which play the same role like the atoms in quantum mechanics \cite{corda}. In this framework, Bekenstein
has found a fundamental result indicating the maximum entropy of the black
hole which is allowed by quantum theory and general theory of  relativity
\cite{bekenstein1} for a given mass and size. The Bekenstein bound puts an upper bound on the
entropy of the system with a finite amount of energy and a given size. This bound is the maximum amount of information required to describe a system by considering its quantum properties \cite{bekenstein1}. If the energy and size of the system is finite, the information required to describe it completely, is finite too.
One of the important consequences of Bekenstein bound is in the physics of information and in computer science when it is connected  with the so-called Bremermann's Limit \cite{brem}. It puts a maximum information-processing rate for a system with
finite size and energy. Another consequence of Bekenstein bound is the derivation
of the field equations of general theory of relativity \cite{jacbson}. There are
some investigations in trying to find some forms of the bound by considering
consistency of the laws of thermodynamics with the general theory of relativity \cite{2005}. In this framework, a generalization of the Bekenstein bound was derived
by Bousso \cite{1999}, conjecturing an entropy bound with its statistical origin which is valid in all space-times consistent with Einstein's equations. This so-called covariant entropy bound reduces to Bekenstein bound in the system of limited self-gravity \cite{1999}. Another attempt in this regard
has been done by Bousso in considering the systems with cosmological horizon which has led to the so-called D-bound \cite{bosso1}. Bousso has derived D-bound for asymptotically non-flat Schwarzschild-de Sitter black hole solution. One can look for D-bound for other solutions which are not asymptotically flat and include  a cosmological apparent horizon.

Surrounded Vaidya black holes, as asymptotically non-flat solutions,  show interesting results under consideration of D-bound which we  intend to study
them in this paper. In fact, Vaidya solution provides
a non-static solution for the Einstein field equations which is a generalization
of the static  Schwarzschild black
hole solution. This solution  depends on the dynamical mass   $m=m(u)$
as a function of the
retarded time coordinate $u$, and an ingoing/outgoing flow $\sigma(u,r)$.
Because of this feature of Vaidya solution, it can be considered as a classical
model for dynamical black hole which is effectively evaporating or accreting.
The process of spherical symmetric gravitational collapse  has also been studied
by applying the Vaidya solution. On the other hand, this solution is a testing
ground for the cosmic censorship conjecture \cite{R1,sg1,ih1,y1}, see also
\cite{t1} for other application.

  Bousso
has considered D-bound and Bekenstein bound for the {\it stationary} Schwarzschild-de
Sitter solution and found that these two bounds are identified for this specific
solution \cite{1999}. The  Vaidya black hole surrounded by cosmological fields
\cite{yaghoub2} is
a generalization of {\it stationary} Schwarzschild-de
Sitter black hole  to the  case of {\it dynamical} black hole embedded in a {\it dynamical} background. Therefore, from entropic point of view, this generalization can account for the dependance of D-bound and Bekenstein bound on the dynamics of black hole and its surrounding fields\footnote{For non-static systems, as well as static  systems, surrounded
by cosmological fields,  D-bound and Bekenstein bound can be implemented
 because both of them are the direct consequence of GSL.
The GSL has been proven for semiclassical quantum  fields  (rapidly changing with time
while falling across a causal horizon) minimally coupled to general relativity
\cite{wall}.
Moreover, the GSL holds
on any causal horizon \cite{tjack}.}. In this paper, motivated by Bousso's work \cite{1999} on a stationary black hole solution,
we show that the identification of generalized D-bound and Bekenstein
bound for \textit{a dynamical black hole solution} surrounded  by some exotic cosmological fields can be considered as a suitable criterion for selecting the cosmological constant-like field as the most viable field  and  ruling out the other fields such as phantom-like and quintessence-like
fields. It is worth to mention that the physical motivation for demanding the identifications of these entropy bounds is that both of them are direct results
of generalized second law of thermodynamics putting an upper bound on the
same matter system.

The organization of the paper is as follows. In section 2,  we review very briefly the D-bound and Bekenstein
bound. In section 3, we introduce the Vaidya black hole solution.    In section 4, we derive D-bound
and Bekenstein bound for surrounded Vaidya solution by cosmological constant-like
field. In sections 5 and 6, these bounds are obtained for surrounded Vaidya
solution by quintessence-like field and phantom-like field, respectively.
At the end, we give a conclusion  in section 8.

\section{Bekenstein Bound and  D-Bound}
In this section, we study the Bekenstein bound and D-bound and summarize
the important results obtained in \cite{bosso1} (for more details see \cite {bekenstein1,bekenstein2,bekenstein3}).

Bekenstein bound is expressed by the following statement: Isolated, stable thermodynamic systems in asymptotically
flat space are constrained by universal entropy bound
\begin{equation}\label{main}
S_{m}\leqslant 2\pi RE,
\end{equation}
where $R$ is the radius of sphere circumscribing   system and $E$ its total energy.
Bekenstein bound has been considered in two forms, empirical and logical.
\begin{itemize}
\item
Empirical form:
All physically reasonable, weakly gravitating matter systems, satisfy the
Bekenstein
bound \cite{bekenstein2,NN}. Some of the systems saturate the bound. For
example, the bound is saturated by Schwarzschild black hole through $S=\pi R^{2}$
and $R=2E$. It seems that the Bekenstein bound is the tightest one for any physical
system.  There are
some controversial examples which claim the violation of Bekenstein bound \cite{page}. However, some of these counter-examples  are shown  not to correctly
include the whole of the gravitating
matter system in $E$ and including them can restor the Bekenstein bound. The rest of  counter-examples also contain controversial matter and excluding
them from $E$ can restor the Bekenstein bound \cite{page2, wald1}.

\end{itemize}
\begin{itemize}
\item
Logical form: Bekenstein has claimed that for weakly gravitating systems,
the bound is a result of generalized second law of thermodynamics (GSL)
\cite{bekenstein3,bekenstein1,bek6,bek7}.
By the Geroch process (a gedankenexperiment), the system is collapsed into
a large black hole. The entropy of the system (black hole) becomes $\Delta
A/4=8\pi RE$. According to  GSL, $\Delta
A/4-S_{m}\geqslant0$ where $S_{m}$ is the entropy of  lost matter system
before the formation of black hole.
There are also some controversial arguments in that whether one can derive Bekenstein
bound by Geroch process considering quantum effects \cite{unruh1,unruh2,pelath}.
However, there is no certain result coming out of these arguments.
\end{itemize}

D-bound is expressed by the following statement: D-bound is a bound on the entropy of matter systems in de Sitter space which is shown to be closely related to the Bekenstein bound in a flat background \cite{bosso1}. The definition of D-bound on matter entropy in de Sitter space is as follows.
Assume an observer located within his apparent cosmological horizon corresponding
to a matter system, in a universe that is asymptotically de Sitter in the future. The observer moves relative to the matter until the matter is located
at his apparent cosmological horizon. He will realize that  crossing out  of the matter from his apparent cosmological horizon is
a thermodynamic process. The entropy of   system after the matter is crossed
out the cosmological horizon is\begin{equation}\label{2}
S_{0}=\frac{A_{0}}{4},
\end{equation}
where $A_{0}$ is the area of cosmological horizon given by\begin{equation}
A_{0}=\pi r_{0}^{2}=\frac{12\pi}{\Lambda}.
\end{equation}
  The entropy of the initial state is the sum of the matter system's entropy $S_{m}$ and a quarter of the apparent cosmological
horizon
\begin{equation}\label{4}
S=S_{m}+\frac{A_{c}}{4}.
\end{equation}
According to the generalized second law of thermodynamics, the observer concludes that the entropy increases. Thus, by comparing equations ($\ref{2}$) and (\ref{4})
we have

\begin{equation}\label{entropy bound}
S_{m}\leqslant \frac{1}{4}(A_{0}-A_{c}),
\end{equation}
which is the D-bound on the matter system in asymptotically de Sitter space.
The D-bound has been derived by Bousso for entropy of the matter systems
in de Sitter space. It is indicated that the D-bound is the same as Bekenstein
bound of the system in this model. Also, Bousso has achieved the same result
for arbitrary dimensions.
In an another example, the  D-bound entropy for the various possible black hole solutions on a 4-dimensional brane
have been considered in \cite{DDD} . It is found that the D-bound entropy for this solution is apparently different from that of obtained for
the 4-dimensional black hole solutions. This difference is considered as the extra loss of information which comes from  the extra dimension, when an extra-dimensional black hole is moved outward the observer's
cosmological horizon. The obtained results there also have been considered,
by adopting the recent Bohr-like approach to black hole quantum physics for the excited black
holes \cite{DDD}.
\section{Surrounded Vaidya Black Hole Solution}
 The metric of   Vaidya black hole  solution surrounded by   cosmological
fields introduced
in \cite{yaghoub1, yaghoub2} is given by
\begin{equation}\label{me}
ds^{2}=-\left(1-\frac{2M(u)}{r}-\frac{N_{s}(u)}{{r}^{{3\omega_s +1}}
}\right)du^2+2\epsilon dudr+r^2d\Omega^2,
\end{equation}
where $M(u)$, $N_{s}(u)$ and $\omega_s$ are black hole dynamical mass, surrounding
field characteristic parameter and equation of state parameter of the surrounding
field, respectively. {As it is mentioned in \cite{yaghoub1}, the metric (6) is a solution to the Einstein field equations
in four dimension for a fluid that is not a perfect fluid in general. However, the \textquotedblleft averaged"
energy-momentum defined as $T^{\mu}_{\nu}=(-\rho,<\mathcal{T}^{i}_j> )$ where $<\mathcal{T}^{i}_j>=\frac{\alpha}{3}\rho_{s}(u,r)\delta^i_j=p_s(u,r)\delta^i_j$,
can be treated as an effective perfect fluid.} 

In contrast to the stationary spacetimes, the local definitions of the various horizons  do not necessarily coincide
with the location of the event horizon for dynamical black holes \cite{ahNiel}. For such dynamical spacetimes,  one is left with the question: ``\textit{For
which surface should one define the black hole area, surface gravity, temperature
or entropy?}''. The canonical choice is to use the event horizon. However,  there are
some evidences that it is the apparent horizon, and not the event horizon, that plays the key
role in the Hawking radiation \cite{ahhaj, ahhis, ahli, ahzho}, see
also \cite{ahVis, ahHay, ahAsht}. This finding has became a key point in hopes to demonstrate the Hawking radiation in the laboratory using the models of analogue gravity \cite{ahLib}. Therefore, we consider  Bekenstein-Hawking entropy
for  apparent horizons associated to the metric  (\ref{me})
with various cosmological fields. Then, we derive D-bound and Bekenstein bound for these backgrounds.

In the following sections, we investigate the D-bound and Bekenstein bound for  Vaidya black hole surrounded by various cosmological fields. Since
the matter source supporting the geometry (\ref{me}) is not a perfect fluid
\cite{yaghoub1, yaghoub2},
we name these fields as quintessence-like, phantom like and cosmological constant-like fields since they possess an average equation of state like the quintessence,
phantom and cosmological constant in the standard model of cosmology. Then, we compare the D-bound
 with Bekenstein
bound  to show that the more cosmological fields are diluted, the more D-bound
and the Bekenstein bound are identified. { The Bekenstein bound is an entropy bound over a matter system
which is saturated for black holes. This bound is the direct result of generalized
second law of thermodynamics (GSL).
On the other hand, D-bound is also is a direct result of GSL which in our model
puts an upper bound on the entropy of the matter system. This upper bound cannot be greater than the Bekenstein bound. Also it should not be less than the
saturated
form of the Bekenstein bound because a black hole possesses a certain entropy. Hence, one can admit
the identification of these two bounds at least for black holes as the local
matter systems.
}

{Here, we deem it necessary to explain a delicate
point. The D-bound entropy is essentially defined for  a local matter system
before and after its local-crossing through  the cosmological apparent horizon of an observer  [20]. But, here we have considered a local matter system surrounded
by a global cosmological field (quintessence or phantom), and applied the D-bound on this combined ``local-global''
system, the global part of which has no any local behavior  in crossing
the cosmological apparent horizon of the observer. How can   this discrepancy
is solved?   To properly
address this point, we may explain as follows. The direct application of D-bound is meaningless for the cosmological  fields because no observer can imagine a ``global'' cosmological field crossing ``locally'' through  his/her cosmological apparent horizon. However, for a combined system of local matter  and global cosmological fields one may still apply the D-bound by ignoring
the direct roles of global  cosmological fields in D-Bound scenario and merely resorting  to their indirect
roles  on the local behaviors of the matter. For a local black hole,  surrounded by cosmological fields,
these indirect
(effective) impacts can be a shift in the locus of black hole's horizon  and the appearance of new horizons. Therefore,  the comparison  of D-bound and
Bekenstein bound for these combined systems are established through considering
merely the local behaviors of the modified horizons, and the identification of these
bounds leads to ruling out the combined system of a black hole surrounded by quintessence or phantom field. }


\section{D-bound and Bekenstein bound for surrounded Vaidya solution by cosmological constant-like field}

\subsection{D-bound}
Considering the  equation of state parameter by $\omega_{c}=-1$
\cite{yaghoub1,yaghoub2},
the metric ($\ref{me}$) becomes
\begin{equation}\label{metricc}
ds^{2}=-\left(1-\frac{2M(u)}{r}-N_{c}(u)r^{2}\right)du^2+2 dudr+r^2d\Omega^{2}_{2},
\end{equation}
where $N_{c}(u)$ is the normalization parameter for the cosmological field surrounding the black hole. This metric describes a black hole surrounded
by cosmological constant-like field. Positive energy condition on the surrounding cosmological field
leads to $N_{c}(u)>0$ \cite{ yaghoub2}. The cosmological background which has
negative surface gravity decreases the gravitational attraction of the black
hole. This repulsive gravitational effect with
the equation of state parameter $\omega_{c}=-1$ makes the cosmological constant
filed as  the most favored candidates for  the dark energy  responsible for the accelerating expansion of the universe \cite{44}.
 The metric ($\ref{metricc}$) indicates the non-trivial effects of the surrounding
cosmological field which differs from Vaidya black hole in an empty background.
The background cosmological field changes the causal structure of the Vaidya black
hole in an empty space. The causal structure change of Vaidya to Vaidya-de Sitter space
is similar to the causal structure change of Schwarzschild to Schwarzschild-de Sitter space
\cite{mallet}.

To derive D-bound for the Vaidya case one needs
the apparent cosmological horizon which will be described  completely in this
section. First, we have to find the  horizons of this solution.
In Ref. \cite{ yaghoub2}, the black hole horizon and the apparent cosmological horizon are obtained for Vaidya solution surrounded by cosmological constant-like field, in detail. There are   black hole and apparent
cosmological horizons, subject to a particular condition   $\Delta(u)=1-27M^{2}(u)N_c(u)>0$,
representing the inner and outer horizons, respectively as \cite{ yaghoub2}
\begin{eqnarray}\label{horizon}
&&r_{AH^{-}}=2M(u)+8M^{3}(u)N_{c}(u)+O(N_{c}^{2}(u)),\\
&&r_{AH^{+}}=\frac{1}{\sqrt{N_{c}(u)}}-M(u)-\frac{3}{2}M^{2}(u)\sqrt{N_{c}(u)}-4M^{3}(u)+O(N_{c}^{\frac{3}{2}}(u)).
\label{h6}\end{eqnarray}
The inner apparent horizon
$r_{AH^{-}}$ is larger than the dynamical schwarzschild radius $r(u)=2M(u)$ and the outer cosmological
apparent
horizon $r_{c}=r_{AH^{+}}$ tends to infinity for $N_{c}(u)\lll 1$. The cosmological field
and the black hole mass have positive contributions to the inner apparent horizon,
whereas the black hole mass has negative contribution to the outer horizon and  pulls the cosmological horizon back towards the center of the black hole. The black hole evaporation leads to shrinking and vanishing of the inner apparent horizon while the outer horizon is tending to reach its  asymptotic value $N_{c}^{-\frac{1}{2}}$.

Now, we apply the Bousso's method like the one defined to some extend in the section 2. We consider an observer inside a system which is circumscribed by a sphere of radius $r_{AH^{+}}$
 $(\ref{h6})$
. Then, we assume that the observer moves away from the matter system (black hole) until he/she observes  that the matter system crosses out  his cosmological apparent
horizon   with radius of $r_{AH^{+}}$.
The GSL claims that the entropy of  final state of this apparent cosmological  horizon in the absence of black hole  is greater
than the entropy of the initial state of this apparent cosmological  horizon
with black hole. The
  final state system circumscribing of sphere with the radius $r_{0}$ has entropy
  $S_{0}=\frac{A_{0}}{4}=\pi r_{0}^{2}$, where $A_{0}$  is the area of the
apparent
  cosmological horizon in the absence of the matter system $(M(u)=0)$ and $r_{0}=r_{c}(M(u)=0)=N_{c}^{-\frac{1}{2}}$. The entropy of
  initial state
  system  is the sum of the matter system (black hole) entropy $S_{m}=S_{AH^{-}}$ and the entropy
  of cosmological horizon $S_{AH^+}$. We can write them as follows
\begin{eqnarray}\label{a1}
&&S_{AH^{-}}=\pi r^{2}_{AH^-}=\pi (4M^2(u)+32M^{4}(u)N_c(u))+O(N^{2}_{c}(u))),\\
&&S_{AH^+}=\pi r^{2}_{AH^+}=\pi \Big(\frac{1}{N_{c}(u)}-2\frac{M(u)}{\sqrt{N_{c}}}-2M^{2}(u)
-5M^{3}(u)\sqrt{N_{c}(u)}-16M^{4}(u)N_{c}
(u))+ O(N^{\frac{3}{2}}_{c}(u)\Big)\label{ss2}
\end{eqnarray}
According to GSL the final entropy $S_{0}=\frac{A_{0}}{4}$ is greater than
initial entropy $S_{AH^{-}}+S_{AH^{+}}$. Thus, using ($\ref{a1}$) and ($\ref{ss2}$)  in $S_{0}\geqslant S_{AH^{-}}+S_{AH^{+}}$, we obtain

\begin{equation}\label{c1}
S_{m}\leqslant  \pi \left( 2\frac{M(u)}{\sqrt{N_{c}}}
+2M^{2}(u)+5M^{3}(u)\sqrt{N_{c}(u)}+16M^{4}(u)N_{c}
(u))\right).
\end{equation}
This is the D-bound for surrounded Vaidya solution by cosmological constant-like field. { When the gravitational radius of the matter system
$r_{g}$ is very smaller
than the cosmological radius $r_{c}$, the system is called \textquotedblleft dilute", i.e. when $r_{g}\ll r_{c}$.
In the Vaidya solution it means that $N_{c}(u)\lll1 $ leading to a large radius for the cosmological horizon.}
   In dilute limit, i.e. $N_{c}(u)\lll1 $   the inequality ($\ref{c1}$) becomes
\begin{equation}\label{limitentropy}
S_{m}\leqslant  2\pi \frac{M(u)}{\sqrt{N_{c}}}.
\end{equation}
The inequality ($\ref{limitentropy}$) puts  an upper bound for
the entropy of the black hole. The normalization parameter for the cosmological field $N_{c}(u)$ in the limit of dilute field makes larger the upper bound
for the black hole entropy but the black hole mass $M(u)$ has opposite role.
One can recognize from inequality ($\ref{c1}$) that,  all terms in RHS  are  positive
or both parameters $M(u)$ and $N(u)$ have positive effects on the upper entropy
of black hole, imposed by D-bound. 
\subsection{Bekenstein bound}
 To derive Bekenstein bound ($\ref{main}$) for surrounded Vaidya solution by cosmological constant-like field we need to know the radius of the sphere
$R$
circumscribing the system and its energy $E$. To find the Bekenstein
bound we will apply the Bousso's method  \cite{bosso1}. For Vaidya black hole surrounded by cosmological field, the energy of the system is not well-defined, due to
the lack of a suitable
asymptotic region. However, there exists a solution which is known as Vaidya black hole solution surrounded by cosmological field which behaves
like  the metric of de Sitter space with cosmological horizon radius $r_{c}$, at large distances. This solution is like  the ``\textit{system's equivalent black hole}'' , and its radius is like the ``\textit{system's gravitational radius}'' $r_{g}$.
The $r_{g}$ for Schwarzschild black hole  equals the twice  energy
of the black hole which is the same as event horizon radius of the black hole. But, for this solution there is some delicate points, as follows. Here the $r_{g}$ is the same
as apparent horizon of the black hole, but it is not the same as  twice
energy of the black
hole. Thus, the corrected $r_{g}$ and cosmological horizon $r_{c}$ are
 \begin{equation}\label{b1}
r_{g}= 2m=r_{AH^{-}}=2M(u)+8M^{3}(u)N_{c}(u)+O(N_{c}^{2}(u)),
\end{equation}
\begin{equation}\label{b2}
r_{c}= r_{AH^{+}}=\frac{1}{\sqrt{N_{c}(u)}}-M(u)-\frac{3}{2}M^{2}(u)\sqrt{N_{c}(u)}-4M^{3}(u)+O(N_{c}^{\frac{3}{2}}(u)).
\end{equation}
The Bekenstein bound for the system's equivalent black hole  with  gravitational
radius $r_{g}$ is written as follows \cite{bosso1}
\begin{equation}\label{b3}
S_{m}\leqslant \pi r_{g}R,
\end{equation}
 where $R$ is  radius of the sphere which circumscribes the system. Here, $R$ is
equal to $r_{c}$. Now, we
 put the equations ($\ref{b1}$) and ($\ref{b2}$) into ($\ref{b3}$). Then, we have
\begin{equation}\label{d1}
S_{m}\leqslant \pi ( \frac{2M(u)}{\sqrt{N_{c}(u)}}+5M^{3}(u)\sqrt{N_{c}(u)} -2M^{2}(u)-8M^{4}(u)-8M^{4}(u)N_{c}(u)-32M^{6}N_{c}(u)+ O(N_{c}^{\frac{3}{2}}(u)).
\end{equation}
We see that   in the  inequality (\ref{d1})  for $N_{c}(u)\lll1$   the first term
 dominates which leads exactly  to the  D-bound. However, in inequality ($\ref{c1}$) for $N_{c}(u)\lll1$
the dominant term  is the first term which is exactly the same as the dominant term in equation
($\ref{d1}$) (i.e $S_{m}\leqslant \pi  \frac{2M(u)}{\sqrt{N_{c}(u)}}$ ). So, Bekenstein
bound and D-bound ($\ref{limitentropy}$) are identified for very diluted surrounding field. For the case of a little less diluted surrounding
field, i.e $N_{c}(u)\ll1$, the Bekenstein bound ($\ref{d1}$) reads
\begin{equation}\label{18}
S_{m}\leqslant \pi ( \frac{2M(u)}{\sqrt{N_{c}(u)}}-2M^{2}(u)-8M^{4}(u)),
\end{equation}
which is a tighter bound than D-bound $S_{m}\leqslant  \pi (\frac{2M(u)}{\sqrt{N_{c}(u)}}+2M^{2}(u))$
derived by this less dilute approximation from
($\ref{c1}$).

{When the field is strong and the system is not dilute, the bounds (\ref{c1}) and
 (\ref{d1})  are not identified. Except for very dilute system limit, the Bekenstein bound  ($\ref{d1}$)
  for the surrounded Vaidya solution by cosmological constant-like field and its D-bound ($\ref{c1}$)
are not the same.} The parameters $N_{c}(u)$ and $M(u)$ always have  positive effects in
the D-bound, but in Bekenstein bound they have both positive and negative
contributions. If negative parts dominate in Bekenstein bound to positive ones,
then RHS in (\ref{18}) becomes negative which is physically meaningless. The requirement
of a positive upper bound in Bekenstein bound puts constraint on  the parameters $N_{c}(u)$ and $M(u)$ in ($\ref{18}$). There is no such
a constraint  for D-bound ($\ref{c1}$) regarding this solution because  RHS in ($\ref{c1}$) is
always positive.
\section{D-bound and Bekenstein bound for surrounded Vaidya black hole by quintessence-like field}
\subsection{D-bound}
Considering the  equation of state parameter $\omega_{q}=-\frac{2}{3}$
\cite{yaghoub1,yaghoub2},
the metric ($\ref{me}$) becomes
\begin{equation}\label{metric14}
ds^{2}=-\left(1-\frac{2M(u)}{r}-N_{q}(u)r\right)du^2+2 dudr+r^2d\Omega^{2}_{2}.
\end{equation}
This metric describes a black hole surrounded by quintessence-like field. Here, $N_{q}(u)$ is the normalization parameter for the quintessence-like field surrounding the black hole. Positive energy condition on the surrounding quintessence-like field
leads to $N_{q}>0$ \cite{ yaghoub2}. According to the metric ($\ref{metric14}$), it  is obvious that the surrounding quintessence-like field has non-trivial
contribution to the metric of Vaidya
black hole. The background quintessence-like field changes the
causal structure of black hole solution in comparison to that of the original Vaidya
black hole in an empty background. An almost similar
effect occurs when one immerses a Schwarzschild black hole in a de Sitter background
 which is asymptotically de Sitter \cite{mallet}. Similar to  Vaidya black hole surrounded by cosmological field, the surface gravity
of the black hole here is also negative and it leads to gravitational repulsion.

  Deriving D-bound for this case is the same as the one which we
derived for Vaidya black hole solution surrounded by cosmological field in
the previous section.  For  $\Delta(u)=1-8M(u)N_{q}(u)>0$,
there is two physical inner and outer apparent horizons \cite{ yaghoub2}.
The locations of two apparent horizons for $\Delta(u)>0$ with small quintessence
normalization parameters ($N_{q}\ll M(u)$) are\begin{equation}
r_{AH^{-}}=2M(u)+4M^{2}(u)N_{q}(u)+O(N_{q}^{2}(u)),
\end{equation}
\begin{equation}
r_{AH^{+}}=\frac{1}{N_{q}(u)}-2M(u)-4M^{2}(u)N_{q}(u)+O(N_{q}^{2}(u)).
\end{equation}
The surrounding quintessence field has contributions both in the  physical inner
horizon $r_{AH^{-}}$ (which is larger than dynamical Schwarzschild radius
$r(u)=2M(u)$)  and the outer  horizon $r_{AH^{+}}$ (cosmological horizon) tending
to infinity for $N_{q}(u)\lll 1$. The quintessence field
and black hole mass have positive contributions for inner apparent horizon. The black hole mass has negative contribution for the outer horizon and it pulls cosmological horizon toward inside. The black hole evaporation leads to shrinking and vanishing of the inner apparent horizon while the outer horizon is tending to its  asymptotic value $N_{q}^{-1}$.

The existence of two apparent horizons guarantees considering D-bound. Here, the entropy of the final state system is $S_{0}=\frac{A_{0}}{4}$, where $A_{0}$ is the area of the cosmological horizon  within which  there is no matter system except the quintessence field. The initial state entropy
is the sum of the black hole entropy $S_{m}=S_{AH^{-}}$ and the cosmological horizon entropy $S_{AH^{+}}$. They are given as,\begin{equation}\label{mm1}
S_{AH^{-}}=\pi r^{2}_{AH^-}= \pi (4M^{2}(u)+16M^{3}(u)N_{q}(u))+O(N_{q}^{2}(u)),
\end{equation}

\begin{equation}\label{nnn1}
S_{AH^+}=\pi r^{2}_{AH^+}=\pi \left(\frac{1}{N_{q}^{2}(u)}-\frac{4M(u)}{N_{q}(u)}-4M^{2}(u)-16M^{3}(u)N_{q}(u))+O(N_{q}^{2}(u)\right).
\end{equation}
According to GSL the final entropy $S_{0}=\frac{A_{0}}{4}$ is greater  than
the initial entropy $S_{AH^{-}}+S_{AH^{+}}$. Thus, by using  ($\ref{mm1}$), ($\ref{nnn1}$)  and  $r_{0}=r_{c}(M(u)=0)=N_{q}^{-1}(u)$ in $S_{0}\geqslant S_{AH^{-}}+S_{AH^{+}} $ we have

\begin{equation}\label{squin}
S_{m}\leqslant \pi\left( \frac{4M(u)}{N_{q}(u)}+4M^{2}(u)+16M^{3}(u)N_{q}(u)\right).
\end{equation}
In the limit of $N_{q}(u)\lll1$ the above inequality becomes
\begin{equation}\label{bek2}
S_{m}\leqslant \pi \frac{4M(u)}{N_{q}(u)}.
\end{equation}
The inequality ($\ref{bek2}$) puts an upper bound for
the entropy of the black hole. As the black hole mass more increases or the
background field more dilutes $N_{q}(u)\lll 1$, the bound becomes more looser.
\subsection{Bekenstein bound}
In this case, the derivation method of Bekenstein bound  ($\ref{main}$) is the same
as the method we used in the previous section for Vaidya solution surrounded
by cosmological field. Regarding inequality ($\ref{b3}$), the radius $R$ of sphere
circumscribing the system and gravitational radius $r_{g}$ are necessary for considering
Bekenstein bound. The gravitational radius $r_{g}$ here is not  twice the energy of the Schwarzschild black hole.  It is not well defined, for the
lack of asymptotic flat region, for a surrounded Vaidya black hole by quintessence
field. However, in this solution $r_{g}$ is the location of the apparent
horizon of the black hole $r_{AH^{-}}$. Also the radius $R$ in this solution is equal to the cosmological apparent horizon $r_{c}= r_{AH^{+}}$. They are
given as follows
\begin{equation}\label{be1}
r_{g}=2m=r_{AH^{-}}=2M(u)+4M^{2}(u)N_{q}(u)+O(N_{q}^{2}(u)),
\end{equation}

\begin{equation}\label{be2}
r_{c}= r_{AH^{+}}=\frac{1}{N_{q}(u)}-2M(u)-4M^{2}(u)N_{q}(u)+O(N_{q}^{2}(u)).
\end{equation}
Now, one can put equations ($\ref{be1}$) and ($\ref{be2}$) into equation ($\ref{b3}$)
to derive Bekenstein bound as
\begin{equation}\label{boundx}
S_{m}\leqslant \pi\left( \frac{2M(u)}{N_{q}(u)} -16M^{3}(u)N_{q}(u))+O(N_{q}^{2}(u)\right).
\end{equation}
Similar to the previous case, we use  $R=r_{c}$ as the radius of sphere which circumscribes the system. In the  Bekenstein bound (\ref{boundx}),
in the limit of very dilute energy $N_{q}(u)\lll1$, the dominant term  is the first term
(i.e.  $S_{m}\leqslant \pi \frac{2M(u)}{N_{q}(u)})$ which is the same Bekenstein bound
for the surrounded Vaidya black hole by a quintessence field, and in this limit of very dilute energy, namely $N_{q}(u)\lll1$, the Bekenstein bound
(\ref{boundx}) is  twice tighter than  the
D-bound ($\ref{bek2}$).  Therefore, for quintessence background the D-bound does not give
the Bekenstein bound.

In this limit, the
normalization parameter $N_{q}(u)$, for $N_{q}(u)\lll1$   makes larger the upper bound for Bekenstein
and the mass of the black hole also does the same job  for large amounts
of mass. As $N_{q}(u)$ increases, the absolute values of the first and
the second terms decrease and increase, respectively on the RHS of ($\ref{boundx}$) and make
tighter the bound. Overall, it turns out that the Bekenstein bound here is tighter than the D-bound. In
the D-bound the mass of black hole has always positive contribution on the upper bound,
but in the Bekenstein bound  (\ref{boundx}) the mass of the black hole may have both positive and negative
contributions.

 \section{D-bound and Bekenstein bound for surrounded Vaidya black hole by phantom-like field}
\subsection{D-bound}
Considering  the equation of state parameter  $\omega_{p}=-\frac{4}{3}$
\cite{yaghoub1,yaghoub2},
the metric ($\ref{me}$) becomes \begin{equation}\label{metricp}
ds^{2}=-\left(1-\frac{2M(u)}{r}-N_{p}(u)r^{3}\right)du^2+2 dudr+r^2d\Omega^{2}_{2},
\end{equation}
which describes a black hole embedded in phantom background. Here, $N_{p}(u)$ is the normalization parameter for the phantom-like field surrounding the black hole. The positive energy condition on the surrounding phantom field
leads to $N_{p}>0$ \cite{ yaghoub2}. According to the metric ($\ref{metricp}$),
the  surrounding phantom field has non-trivial effect on the Vaidya black hole and its causal structure.  In this case, like  the previous cases,  the phantom-like background field causes a negative surface gravity which leads to the gravitational repulsion\cite{44}.

In deriving D-bound for this case, we are interested in the solutions
with two apparent horizons,  one of them is black hole apparent horizon and the other one plays the role of cosmological horizon $r_{c}$. These solutions
are obtained by  the condition $\Delta (u)=1-\frac{2048}{27}M^{3}(u)N_{p}(u)>0$
as \cite{yaghoub2}. They are as follows
\begin{equation}\label{app1}
r_{AH^{-}}=2M(u)+16M^{4}(u)N_{p}(u)+O(N_{p}^{2}(u)),
\end{equation}
\begin{equation}\label{app2}
 r_{AH^{+}}=\frac{1}{N_{p}^{\frac{1}{3}}(u)}-\frac{2}{3}M(u)-\frac{8}{9}M^{2}(u)N_{p}^{\frac{1}{3}}(u)
-\frac{160}{81}M^{3}(u)N_{p}^{\frac{2}{3}}(u)-\frac{16}{3}M^{4}(u)N_{p}(u)+O(N_{p}^{\frac{4}{3}}(u)),
\end{equation}
where $r_{AH^{-}}$ and $r_{AH^{+}}$ represent the inner and outer physical apparent horizons, respectively.
Thus,  the  black hole in a phantom background
posses an inner horizon larger than dynamical
Schwarzschild radius (apparent horizon) $r(u)=2M(u)$ and an outer horizon,
which is cosmological apparent horizon, blows up for $N_{p}\lll 1$. Regarding equations ($\ref{app1}$) and ($\ref{app2}$), one can realize that
the phantom field makes larger the inner apparent horizon and the black hole
mass makes larger the outer horizon. The black hole evaporation process shrinks
the inner apparent horizon while the outer one closes  to its asymptotic value
$N_{p}^{-1/3}$.

  The existence of two apparent horizons guarantees considering D-bound. Here, the entropy of the final state  is $S_{0}=\frac{A_{0}}{4}$, where $A_{0}$ is the area of the cosmological horizon in the absence of matter system  except the phantom field. The initial state entropy
is the sum of the black hole entropy $S_{m}=S_{AH^{-}}$ and the cosmological horizon entropy $S_{AH^{+}}$ given as
\begin{equation}\label{entp1}
S_{AH^{-}}=\pi r^{2}_{AH^-}= \pi (4M^{2}(u)+64M^{5}(u)N_{p}(u)) + O(N_{p}^{2}(u)),
\end{equation}
\begin{eqnarray}\label{entp2}
&&S_{AH^{+}}=\pi r_{AH^{+}}^{2}=\\ \nonumber
&& \pi (\frac{1}{N_{p}^{\frac{2}{3}}(u)}-\frac{4}{3}\frac{M(u)}{N_{p}^{\frac{1}{3}}(u)}-\frac{4}{3}M^{2}(u)
-\frac{224}{81}M^{3}(u)N_{p}^{\frac{1}{3}}(u)- \frac{1760}{243}M^{4}(u)N_{p}^{\frac{2}{3}}(u)-\frac{64}{3}M^{5}(u)N_{p}(u))+
O(N_{p}^{\frac{4}{3}}(u)).
\end{eqnarray}
  According to GLS and by using ($\ref{entp1}$), ($\ref{entp2}$) and  $r_{0}=r_{c}(M(u)=0)=N_{p}^{-\frac{1}{3}}$  in $S_{0}\geqslant S_{AH^{-}}+S_{AH^{+}} $, one can derive the D-bound for this solution as

\begin{equation}\label{hamed}
S_{m}\leqslant \pi (\frac{4}{3}\frac{M(u)}{N_{p}^{\frac{1}{3}}(u)}+\frac{4}{3}M^{2}(u)
+\frac{224}{81}M^{3}(u)N_{p}^{\frac{1}{3}}(u)+ \frac{1760}{243}M^{4}(u)N_{p}^{\frac{2}{3}}(u)+\frac{64}{3}M^{5}(u)N_{p}(u)).
\end{equation}
 In the limit $N_{p}\lll1$,  the  D-bound ($\ref{hamed}$) becomes
\begin{equation}\label{bound1}
S_{m}\leqslant  \frac{4}{3}\pi \frac{M(u)}{N_{p}^{\frac{1}{3}}(u)}.
\end{equation}
The inequality ($\ref{bound1}$) represents an upper bound for
the entropy of the black hole. The upper bound
 entropy
becomes larger for large black hole mass $M(u)$   and small
normalization parameter  $N_{p}(u)$.

\subsection{Bekenstein bound}

 The gravitational radius and the outer cosmological
apparent horizon in this case are obtained as
\begin{equation}\label{e1}
r_{g}=2m=r_{AH^{-}}=2M(u)+16M^{4}(u)N_{p}(u)+O(N_{p}^{2}(u)),
\end{equation}
\begin{equation}\label{e2}
r_{c}= r_{AH^{+}}=\frac{1}{N_{p}^{\frac{1}{3}}(u)}-\frac{2}{3}M(u)-\frac{8}{9}M^{2}(u)N_{p}^{\frac{1}{3}}(u)
-\frac{160}{81}M^{3}(u)N_{p}^{\frac{2}{3}}(u)-\frac{16}{3}M^{4}(u)N_{p}(u)+O(N_{p}^{\frac{4}{3}}(u)).
\end{equation}
We can put equations ($\ref{e1}$) and ($\ref{e2}$) into equation ($\ref{b3}$)
to derive the Bekenstein bound for $N_{p}\lll1$ as \begin{equation}\label{m1}
S_{m}\leqslant \pi(\frac{2M(u)}{N_{p}^{\frac{1}{3}}(u)}-\frac{4}{3}M^{2}(u)).
\end{equation}
 In the Bekenstein bound ($\ref{b3}$) we put  $R=r_{c}$ as the radius of sphere circumscribing the system. In this case, if $\frac{1}{N_{p}^{\frac{1}{3}}(u)}< 4M(u)$
the Bekenstein bound ($\ref{m1}$) will be tighter than the D-bound ($\ref{bound1}$).
But, we know that $\Delta (u)=1-\frac{2048}{27}M^{3}(u)N_{p}(u)>0$  which
leads to $\frac{1}{N_{p}^{\frac{1}{3}}(u)}>
4M(u)$ gives two real solutions  as the physical apparent horizons which are necessary
for considering the D-bound. The other amounts of $\Delta$ (i.e $\Delta(u)\leqslant 0$) which lead to $\frac{1}{N_{p}^{\frac{1}{3}}(u)}<
4M(u)$ cannot give two physical apparent horizons as solutions. So, the Bekenstein bound here, cannot
be tighter than the D-bound. Therefore, if $\frac{1}{N_{p}^{\frac{1}{3}}(u)}> 4M(u)$ the D-bound will be tighter than the Bekenstein bound. However, for $\frac{1}{N_{p}^{\frac{1}{3}}(u)}= 4M(u)$  there is no D-bound because we
have not two physical apparent horizons.

\section{Conclusions and results}
{The D-bound entropy is essentially defined for  a matter system
which crossing  a cosmological apparent horizon of the observer who is near the
matter system. Hence the definition of D-Bound is based on the existence of a  cosmological horizon  far from the local matter system and
their corresponding entropies. Since the cosmological field fills
whole of the system and defines the cosmological
horizon,
the global structure of the system  is  important. In principle, we use each
of the  D-bound and Bekenstein bound to find  a constraint
on the entropy of the matter system.  The constraint on the cosmological field surrounding the matter system is a consequence of the  \textquotedblleft \textit{identification}" of these two bounds. This demand for the identification of these two bounds is based on the
GSL.
} We have derived the D-bound for the Vaidya solutions surrounded by cosmological fields and indicated that for the one particular  solution the D-bound is the same as the Bekenstein bound in dilute systems. 
{Before representing our conclusions and results for the D-bound and Bekenstein
bound defined for the matter system's equivalent black hole, it is worth
to mention that  the Vaidya geometry studied in this work also admits naked singularity type solutions \cite{nake}. The possibility of the formation of naked singularities
for the solution (6) is addressed in \cite{yaghoub1}. In summary, if the discriminant
(equation (25) in \cite{yaghoub1}) for the null geodesics admits a positive real root, then the solution describes a naked singularity and consequently provides a counterexample for the Penrose's cosmic censorship conjecture
\cite{pen}.In contrast to black holes, naked singularities  don't have  any apparent
or event horizon, and then one cannot adjust a suitable radius circumscribing
these systems. As a consequence, it makes difficult to define a Bekenstein bound for such systems. The same argument applies to the case of D-bound that is defined based on the exitance of cosmological horizon  and  a matter system which is called by Bousso as \textit{system's equivalent black hole} possessing a  radius named as the \textit{system's gravitational radius $r_g$} \cite{bosso1}. For  the case of a naked singularity, the definition of a suitable gravitational radius $r_g$ is not trivial.}

The results obtained in the present work are as follows:
\begin{itemize}
\item
The D-bound   for surrounded Vaidya solution by a cosmological constant-like field is the same as Bekenstein bound  in the dilute system limit. As the background field becomes more considerable, the equality
of D-bound and Bekenstein bound is more ruined. In the case of dilute cosmological constant-like field, the  contribution
of background field in the metric is $N_{c}r^{2}$   which leads to  the equality of D-bound
with Bekenstein bound.
 \end{itemize}

\begin{itemize}
\item
The D-bound  for the  Vaidya black hole surrounded by
an exotic quintessence-like field is the same
as Bekenstein bound in light background field, except for a constant coefficient 2.   Since the contribution of  background field
in the metric is $N_{q}r$,  which is weaker than the case of cosmological constant
like field $N_{c}r^{2}$ at $r>1$, the D-bound does not   again coincide with the Bekenstein bound, even  in the light background systems.  In light quintessence-like background field  the D-bound
is looser than the Bekenstein bound.   \end{itemize}
\begin{itemize}
\item

The D-bound  for the  Vaidya black hole surrounded
by an exotic phantom-like field is not the same
as Bekenstein bound in dilute phantom  background field.   Since the contribution of  background field
in the metric is $N_{p}r^{3}$,  which is stronger than the case of cosmological constant-like field $N_{c}r^{2}$ at $r>1$, the D-bound does not again coincide with the Bekenstein bound, even  in the light background systems. The D-bound
is tighter than the Bekenstein bound for the light phantom background.
\end{itemize}

The conclusions are as follows. The dynamical background fields, possessing cosmological horizons,
play the role of a repulsion force like  the case of a cosmological
constant which manifests
itself   in the metric as $r^{2}$. For
this repulsion force,  the D-bound is identified with the Bekenstein bound
in dilute systems. Any deviation from  $r^{2}$ term corresponding to the  quintessence and phantom fields
with contributions as  $r$ and $r^{3}$ terms having less  and
more repulsion forces than that of the cosmological constant leads to D-bounds
looser and tighter than the Bekenstein bound, respectively.  At the end, it is worth mentioning that D-bound and Bekenstein
bound are the direct consequence of GSL. Therefore, we conclude that both of them
should lead to the same entropy bound  imposing on a certain matter
system. This conclusion leads to one possible option as follows:
\begin{itemize}
\item \textbf{Cosmological constant-like field viability:} The cosmological constant-like field has a reasonable behaviour, among two other
cosmological fields, namely the quintessence and phantom, regarding the identification
of D-bound and Bekenstein bound for light systems. It seems that by implementation
of a thermodynamical criterion, namely the \textit{identification of the D-bound and Bekenstein
bound},  on the Vaidya
black hole solution surrounded  by  cosmological fields, one
may exclude the quintessence and phantom fields and just keep the cosmological
constant as the   single field for which D-bound and Bekenstein
bound are exactly identified.\footnote{The violation of second law of thermodynamics by quintessence
and
phantom fields
which  represents
 their
un-physical behaviors in many ways, has been discussed in \cite{ph1,ph2,ph3,ph4,ph5,ph6,ph7}.}
\end{itemize}
{We have studied the same thermodynamical criterion  on the other known
dynamical black hole solutions surrounded  by cosmological fields and observed
that the cosmological constant is preferred as the viable cosmological
field in comparison to the other known cosmological fields  \cite{ub,MC}.
In  \cite{MC},
 the hypothesis of \textquotedblleft\textit{ D-bound-Bekenstein bound identification}" has been applied for the McVittie solution surrounded by cosmological fields in the dilute
limit and it has been indicated that this criterion is only true for cosmological
constant field as a candidate for the dark energy, and the other cosmological fields
such as phantom and quintessence do not satisfy this criterion.  Also in \cite{ub} the same criterion is applied for  the Universe system and the Universe-Black hole system,
which is dominated by the quintessence, phantom or cosmological constant fields. By using the entopic considerations, it turns out that for both systems,
the
cosmological constant field is the only viable field among the others.} 


\begin{thebibliography}{99}

\bibitem{CP}P. J. Steinhardt , in Critical Problems in Physics, edited
by V. L. Fitch and Dr. R. Marlow (Princeton University
Press, Princeton, N. J., 1997);
I. Zlatev, L. M. Wang, and P. J. Steinhardt, Phys. Rev.
Lett. 82 (1999),  896; P. J. Steinhardt, L. M. Wang, and
I. Zlatev, Phys. Rev. D 59  (1999).
\bibitem{CCP}S. Weinberg,  Rev. Mod. Phys.  61, 1 (1989).
\bibitem{Caldwell}R. R. Caldwell, R. Dave, and P. J. Steinhardt,  Phys. Rev. Lett. 80 (1998),
1582.
\bibitem{fujii}Y. Fujii, Phys. Rev. D 26 (1982), 2580.
\bibitem{ford} L. H. Ford, Phys. Rev. D
35 (1987), 2339.
\bibitem{wett} C. Wetterich, Nucl. Phys. B 302
(1988), 668.
\bibitem{ratra} B. Ratra and P. J. E. Peebles,  Phys. Rev. D 37 (1988), 3406.
\bibitem{carroll12}S. M. Carroll, M. Hoffman, and M. Trodden,  Phys. Rev. D 68 (2003), 023509.
\bibitem{singh12} P. Singh, M. Sami, and N. Dadhich,
Phys. Rev. D 68 (2003), 023522.
\bibitem{sami12} M. Sami and A. Toporensky, Mod. Phys.
Lett. A 19 (2004), 1509.
\bibitem{bekenstein3}J. D. Bekenstein, Phys. Rev. D 7, 2333 (1973).
\bibitem{moon1}J. D. Bekenstein, Lett. Nuovo Cim. 11, 467 (1974).
\bibitem{hawking1}S.W. Hawking, Commun. Math. Phys. 43, 199 (1975).

\bibitem{corda}C. Corda, Class. Quantum Grav. 32, 195007 (2015).
\bibitem{bekenstein1}J. D. Bekenstein, Phys. Rev. D 23, 287 (1981).
\bibitem{brem}H. J. Bremermann, ``{\it Optimization through evolution and recombination}'', In: Self-Organizing systems
1962, edited M.C. Yovitts et al., Spartan Books, Washington, D.C. pp. 93 -106 (1962).
\bibitem{jacbson}T. Jacbson, Phys. Rev. Lett. 75, 1260 (1995).
\bibitem{2005}J. D. Bekenstein, Cont. Phys. 45 (1), 31 (2005).
\bibitem{1999}R. Bousso, JHEP 07, 004 (1999).
\bibitem{bosso1}R. Bousso, JHEP 0104,  035 (2001).
\bibitem{R1}R. Penrose, Riv. Nuovo Cimento 1, 252 (1969).
\bibitem{sg1}S.G. Ghosh, N. Dadhich, Phys. Rev. D 64, 047501 (2001).
\bibitem{ih1}I.H. Dwivedi, P.S. Joshi, Class. Quant. Grav. 6(11), 1599 (1989).
\bibitem{y1}Y. Kuroda, Y. Prog, Theor. Phys. 72(1), 63 (1984).
\bibitem{t1}T. Harko, Phys. Rev. D 68, 064005 (2003).
\bibitem{yaghoub2}Y. Heydarzade, F. Darabi, Eur.Phys.J. C78, 342 (2018).
\bibitem{bekenstein2}J. D. Bekenstein,
Phys. Rev. D 30, 1669 (1984).
\bibitem{NN}M. Schiffer and J. D. Bekenstein, Phys. Rev. D 39, 1109 (1989).
\bibitem{page}D. N. Page, Subsystem entropy exceeding Bekenstein's bound (2000), hep-th/0007237.
\bibitem{page2} J. D. Bekenstein, On Page's examples challenging the entropy bound (2000), gr-qc/0006003.
\bibitem{wald1}R. M. Wald,      Living Rev. Rel.4, 6 (2001).
\bibitem{bek6}J. D. Bekenstein, Nuovo Cim. Lett. 4, 737 (1972).
\bibitem{bek7}J. D. Bekenstein, Phys.
Rev. D 9, 3292 (1974).
\bibitem{unruh1}W. G. Unruh and R. M. Wald, Phys. Rev. D 25, 942 (1982).
\bibitem{unruh2}W. G. Unruh and R. M. Wald, Phys. Rev. D 27, 2271 (1983).
\bibitem{pelath}M. A. Pelath and R. M. Wald, Phys. Rev. D 60, 104009 (1999).
\bibitem{DDD}Y. Heydarzade, H. Hadi, C. Corda, F. Darabi, Phys. Lett. B 776, 457 (2018).
\bibitem{yaghoub1}Y. Heydarzade, F. Darabi, Eur. Phys. J. C78, 582 (2018).
\bibitem{ahNiel}A. B. Nielsen, Black holes as local horizons, arXiv:0711.0313.
\bibitem{ahhaj}P. Hajicek, Phys. Rev. D 36, 1065 (1987).
\bibitem{ahhis}W. A. Hiscock, Phys. Rev. D 40, 1336 (1989).
\bibitem{ahli}X. Liou, W. Liou, Int. J. Theor. Phys. 49. 5, 1088 (2010).
\bibitem{ahzho}S. Zhou, W. Liu, Mod. Phys. Lett. A  24. 26, 2099 (2009).
\bibitem{ahVis}M. Visser, Int. J. Mod. Phys. D 12, 649 (2003).
\bibitem{ahHay} S.A. Hayward, Phys. Rev. D 49, 6467 (1994).
\bibitem{ahAsht} A. Ashtekar, B. Krishnan, Living
Rev. Rel. 7, 10 (2004).
\bibitem{ahLib} C. Barcelo, S. Liberati, M. Visser, Living Rev. Rel. 8, 12 (2005).
\bibitem{44}A. Vikman, Phys. Rev. D 71, 023515 (2005).
\bibitem{mallet}R. L. Mallett, Phys. Rev. D 31, 2, 416 (1985).
\bibitem{bonnor}W. B. Bonnor, P. C. Vaidya, Gen. Rel. Grav. I, 2, 127 (1970).
\bibitem{1}V. V. Kiselev,  Class. Quant. Grav. 20, 1187–1197 (2003).
\bibitem{mc} G. C. McVittie,  Mon. Not. R. Astr. Soc. 93, 325
(1933).
\bibitem{wall}A. C. Wall, Phys. Rev. D 85,
104049 (2012).
\bibitem{ph1}I. Brevik, S. Nojiri, S. D. Odintsov,  L. Vanzo, Phys. Rev. D 70, 043520 (2004).
\bibitem{ph2}P. F. Gonzalez-Diaz, C. L. Siguenza, Nucl. Phys. B 697, 363 (2004).
\bibitem{ph3}D. H. Hsu, A. Jenskins, M. B. Wise, Phys. Lett. B 597, 270 (2004).
\bibitem{ph4}J. A. S. Lima, J. S. Alcaniz,  Phys. Lett. B 600, 191 (2004).
\bibitem{ph5}H. Mosheni Sadjadi,  Phys. Rev. D 73, 0635325 (2006).
\bibitem{ph6}S. Nojiri, S. D. Odintsov, Phys. Rev. D 70, 103522 (2004).
\bibitem{ph7}S. Nojiri, S. D. Odintsov,   Phys. Lett. B 595, 1 (2004).
\bibitem{tjack}T. Jacobson and R. Parentani,  Found. Phys. 33, 323 (2003).
\bibitem{ub}H. Hadi, F. Darabi, Y. Heydarzade, EPL (Europhysics Letters)
131 (5), 59001 (2020)
\bibitem{MC}H. Hadi, Y. Heydarzade, F. Darabi, K. Atazadeh, The European
Physical Journal Plus 135 (7), 1-14 (2020).
\bibitem{nake}P. Rudra, R. Biswas, U. Debnath, Astrophys. Space. Sci
354, 597 (2014).
\bibitem{pen}R. Penrose, Riv. Nuovo Cimento 1, 252 (1969).
\end{thebibliography}
\end{document}